\documentclass{article}
\pdfoutput=1

\usepackage[margin = 3.1cm]{geometry}
\usepackage{amsmath}
\usepackage{amsthm}

\usepackage{centernot}
\usepackage{braket}

\usepackage[bitstream-charter,cal=cmcal]{mathdesign}
\usepackage[T1]{fontenc}
\usepackage[normalem]{ulem}

\usepackage{xcolor}
\definecolor{dullmagenta}{rgb}{0.4,0,0.4}   
\definecolor{darkblue}{rgb}{0,0,0.4}
\usepackage[bookmarks,colorlinks,breaklinks]{hyperref}  
\hypersetup{linkcolor=red,citecolor=blue,filecolor=darkblue,urlcolor=darkblue} 

\usepackage{verbatim}

\newcommand*{\Laws}{{\normalfont [\text{Laws}]}}
\newcommand*{\ClassicalLaws}{{\normalfont[\text{Classical Laws}]}}
\newcommand*{\Protocol}{{\normalfont[\text{Protocol}]}}
\newcommand*{\AbsPro}{{\normalfont[\text{Abstract Protocol}]}}
\newcommand*{\PhysPro}{{\normalfont[\text{Physical Protocol}]}}
\newcommand*{\Secrecy}{{\normalfont[\text{Secrecy}]}}
\newcommand*{\Leakage}{{\normalfont[\text{Leakage}]}}
\newcommand*{\NoiseThreshold}{{\normalfont[\text{AccurateControl}]}}

\usepackage[backend=biber,sorting=none,style=phys-jmr,biblabel=brackets,backref=false,bibencoding=utf8,maxnames=10,eprint=true]{biblatex}
\makeatletter
\pretocmd{\blx@head@bibintoc}{\phantomsection}{}{\ddt}
\makeatother
\DefineBibliographyStrings{english}{%
  backrefpage = {page},
  backrefpages = {pages},
  phdthesis = {PhD dissertation}
}
\usepackage{titlesec}

\titleformat*{\section}{\normalsize \bfseries}
\titleformat{\subsection}[runin]{\normalsize\bfseries}{\thesubsection. }{0.0em}{}[]
\titleformat*{\subsubsection}{\bfseries}
\titleformat*{\paragraph}{\normalsize\bfseries}
\titleformat*{\subparagraph}{\large\bfseries}

\titlespacing*\section{0pt}{12pt plus 4pt minus 2pt}{2pt plus 2pt minus 2pt}
\titlespacing*\subsection{0pt}{8pt plus 4pt minus 2pt}{4pt plus 2pt minus 2pt}

\title{{\Large Are quantum cryptographic security claims vacuous?}}

\author{Joseph M.\ Renes and Renato Renner\\
\normalsize Institute for Theoretical Physics, ETH Z\"urich}

\date{}

\begin{document}
\renewcommand{\abstractname}{\vspace{-\baselineskip}}

\maketitle

\vspace{-.5cm}
\begin{abstract}
A central claim in quantum cryptography is that secrecy can be proved rigorously, based on the assumption that the relevant information-processing systems obey the laws of quantum physics.  This claim has recently been challenged by Bernstein (\href{http://arxiv.org/abs/1803.04520}{arXiv:1803.04520}). He argues that the laws of physics may also entail an unavoidable leakage of any classical information encoded in physical carriers. The security claim of quantum key distribution would then be vacuous, as the computation of the final secret key would leak its value.  However, as we explain in this short note, Bernstein's reasoning is based on a too ``classical'' understanding of physics. It follows from known theorems about fault-tolerant quantum computation that quantum physics avoids his conclusion.
\end{abstract}
\vspace{0cm}

\section{Criticism of the foundations of quantum cryptography}

The security of Quantum Key Distribution (QKD)~\cite{bennett_quantum_1984,ekert_quantum_1991} can be proved based on the assumption that the devices used to process information obey the laws of quantum theory~\cite{mayers_quantum_1996,shor_simple_2000,christandl_generic_2004,ben-or_universal_2005,renner_security_2005} (for reviews see~\cite{scarani_security_2009,portmann_cryptographic_2014}).\footnote{\label{ftn_correctcomplete}More precisely, one must assume correctness and completeness of quantum theory. \emph{Completeness}  means that it captures all physically accessible information, i.e., that there are no ``hidden variables'' that do not obey the laws of the theory. We refer to~\cite{ColRen11} for a discussion as well as an argument showing that, under a rather general assumption on the causal structure of spacetime, correctness of quantum theory implies its completeness.}  Denoting this assumption by $\Laws$, and writing $\Protocol$ for any QKD protocol, say BB84~\cite{bennett_quantum_1984} or E91~\cite{ekert_quantum_1991}, one may phrase this claim as an implication,\footnote{QKD protocols also require communication over an authentic channel. In this note we will always take the availability of such a channel as being included in assumption~$\Protocol$.} 
\begin{align} \label{eq_securityproof}
 \Laws \wedge \Protocol \quad \implies \quad \Secrecy\, .
\end{align}
Here $\Secrecy$ is the statement that the generated key is, to good approximation,\footnote{\label{ftn_epsilon}The key is allowed to deviate from perfect secrecy with a small probability~$\varepsilon > 0$; cf.\ \cite{portmann_cryptographic_2014} for a precise statement.} uniformly random and independent of all information gathered by an adversary Eve, who may have access to everything except for the labs of the communicating parties Alice and Bob.

This claim has been challenged repeatedly (see \cite{renner_reply_2012} for references). 
Here we reply to a rather fundamental criticism by Bernstein~\cite{bernstein_is_2018}. If correct, Bernstein's argument would imply that all existing theorems that formalize the security statement~\eqref{eq_securityproof}, even if technically sound, are vacuous. 

\subsection{Bernstein's claim.} Bernstein starts with the observation that the description of a quantum-crypto\-gra\-phic protocol is usually provided on an abstract level that is largely independent of the physics of the actual information carriers. For instance, the protocol demands that Alice and Bob compute the outcome of a certain function applied to their own data, but apart from specifying the function, the protocol is silent about the physics of this computation. It leaves open how the data is encoded in physical information carriers, and how the carriers are stored or made to interact.  Instead of~\eqref{eq_securityproof}, it is thus more accurate to say that current security proofs establish the implication
\begin{align} \label{eq_securityproofabstract}
 \Laws \wedge \AbsPro \quad \implies \quad \Secrecy\, ,
\end{align}
where  $\AbsPro$ stands for a specification of the QKD protocol on the abstract level of quantum information theory.

Conversely, since protocols need to be executed on actual physical devices, to claim security of a QKD scheme one would need to prove the implication 
\begin{align} \label{eq_securityproofphysical}
 \Laws \wedge \PhysPro \quad \implies \quad \Secrecy\, ,
\end{align}
where $\PhysPro$ is a full specification of the physical actions to be carried out by the communicating parties. The gap between $\PhysPro$ and $\AbsPro$ is the crux of Bernstein's criticism. His claim is  that, even if~\eqref{eq_securityproofabstract} were correct for a given $\AbsPro$, \eqref{eq_securityproofphysical} may not hold for any $\PhysPro$. Put differently, Bernstein's claim is that the translation of $\AbsPro$ usually found in the QKD literature to $\PhysPro$ may be impossible.\footnote{\label{ftn_theoreticalQKD}This argument is independent of the problem of \emph{``quantum hacking attacks''}, which exploit design flaws in actual implementations, such as the use of devices that do not  adhere to the specifications of $\PhysPro$. Bernstein's argument is aimed at ``theoretical QKD'', excluding these attacks by assumption (see also \cite{scarani_black_2014} in this vein).} The security of QKD would then be inconsistent with the very laws of physics. 

\subsection{Bernstein's argument.} \label{sec_BernsteinArgument} To establish this claim, Bernstein argues that, due to the laws of physics, information unavoidably leaks to the environment when Alice and Bob process secret information with physical systems, such as electronic devices. 
He writes~\cite[Page~4]{bernstein_is_2018}, 
\begin{quote}
  \emph{In light of Maxwell's equations, how can one justify the notion that Eve is unable to observe secret electromagnetic actions by Alice and Bob?} 
  \end{quote}    
He also describes a concrete attack, similar to van Eck phreaking~\cite{van_eck_electromagnetic_1985}, in which Eve builds a large array of radio receivers to measure the electromagnetic field around Alice and Bob. He notes that physical shielding, e.g., with a Faraday cage, merely applies some scrambling to the information leaked to the environment, and hence does not resolve this problem. 
Thus, there would be no way to process information secretly, leading to the claim
\begin{align} \label{eq_leakage}
  \forall \PhysPro:\quad \Laws \wedge \PhysPro \quad \implies \quad \Leakage \, .
\end{align}
Here $\Leakage$ is the assertion that the value of the key generated by the protocol leaks to the environment, and thus becomes accessible to Eve. 

Since $\Secrecy$ and $\Leakage$ obviously exclude each other, implications~\eqref{eq_securityproofphysical} and~\eqref{eq_leakage} are contradictory.  Based on this, Bernstein argues~\eqref{eq_securityproofphysical} cannot be satisfied by any $\PhysPro$, and therefore the security claim of QKD is ultimately vacuous. 

\section{Refutation of the criticism}

Clearly, Bernstein's conclusion hinges on~\eqref{eq_leakage}. We argue that \eqref{eq_leakage} is false and, by describing physical actions Alice and Bob can take to ensure that their information is processed secretly, that \eqref{eq_securityproofphysical} can indeed be established. However, for the description of these actions, it is again crucial to use quantum theory --- secrecy cannot be established within classical physics.

\subsection{Leakage in classical physics.} 
We have no counterargument to Bernstein's claim \eqref{eq_leakage} under the assumption that physical information carriers obey purely classical laws, such as Maxwell's equations. Indeed, any such carrier would necessarily be accompanied by electric, magnetic, or gravitational fields, whose changes could, in principle, be detected at distant locations.\footnote{Note that this also implies Alice and Bob may be able to detect Eve's eavesdropping, but we do not consider this further.}
A correct variant of Bernstein's  claim~\eqref{eq_leakage} would thus be
\begin{align} \label{eq_classicalleakage}
  \forall \PhysPro :\quad\ClassicalLaws \wedge \PhysPro \quad \implies \quad \Leakage \,,
\end{align}
where $\ClassicalLaws$ refers to the assumption that all carriers of classical information (such as the raw key obtained by Alice and Bob) are described by the classical laws of physics.\footnote{It would of course not make sense to require that the carriers of quantum information be described by classical physics.}  

However, $\Laws$ as used in \eqref{eq_securityproofphysical} refers to quantum theory. 
Importantly, although sometimes described as a special case of the quantum laws, the classical laws of physics are not implied by them, i.e., 
\begin{align} \label{eq_quantumclassical}
  \Laws \quad \centernot \implies \quad \ClassicalLaws\,.
\end{align}
In particular, \eqref{eq_classicalleakage} does not rule out~\eqref{eq_securityproofphysical}. That is, if one replaces Bernstein's claim~\eqref{eq_leakage} by the  ``corrected'' version~\eqref{eq_classicalleakage} then the contradiction discussed at the end of Section~\ref{sec_BernsteinArgument} no longer arises.

\subsection{Leakage in quantum physics.} 

Now we explain why the claimed implication~\eqref{eq_leakage} does not hold when $\Laws$ refer to quantum physics.\footnote{Bernstein criticizes, rightly, that $\Laws$ is often not made precise.  Detailed accounts of the formalism of quantum information theory and the physical assumptions that underly it can  be found in, e.g., \cite[Section~2.2]{nielsen_quantum_2010} or \cite{holevo_probabilistic_2011}.} 
More precisely, we argue that there exist ways to process classical information locally without leakage. Since $\Laws$ asserts that all information-processing devices obey the laws of quantum physics, we naturally have to assume that \emph{all} information, including classical information, is represented by the states of a quantum system. The idea is then to exploit the fact that quantum theory imposes a relation between leakage of information and its disturbance. 

The relation between leakage and disturbance is an instance of the uncertainty principle. 
It can be very simply illustrated as follows. Consider a quantum system $Q$ which can store one bit of classical information $X$, encoded into orthogonal basis states $\ket{b_x}$. 
For instance, the two basis states could be two different electronic states of a trapped ion. 
Leakage of the classical bit, e.g., via electromagnetic radiation, corresponds to an operation that copies $X$  to another system, call it $Q'$.
This can be modeled by a \textsc{cnot} gate from $Q$ and $Q'$, the control on $Q$ and the target on $Q'$, with $Q'$ initially prepared in the $\ket{b_0}$ state. In a classical world, the presence of the \textsc{cnot} gate could not be noticed by a party with access to $Q$ only. However, quantum theory asserts that the \textsc{cnot} gate will in general affect the state of system~$Q$ and is thus detectable in principle.  Specifically, to test the system for leakage, $Q$ may be prepared in a superposition state $\ket{+}=\ket{b_0}+\ket{b_1}$ (ignoring normalization). 
The \textsc{cnot} gate will entangle $Q$ and $Q'$, producing $\ket{\Psi}=\ket{b_0}_Q\ket{b_0}_{Q'}+\ket{b_1}_Q\ket{b_1}_{Q'}$. 
The marginal state of $Q$, ignoring system $Q'$, is then just the maximally-mixed density operator, an equal mixture (not superposition) of $\ket{+}$ and $\ket{-}=\ket{b_0}-\ket{b_1}$. 
Thus, the phase of the superposition has been randomized, or, in other words, $Q$ has been subjected to a phase error. 
The observation of a phase error implies that leakage must have occurred.

Although the leakage mechanism in this illustrating example is very specific,  there is a general tradeoff between information gain and disturbance in quantum mechanics. Leakage will always lead to disturbance, no matter the details of the channel via which it leaks.  To phrase this tradeoff on the required level of generality, we describe processes as Trace-Preserving Completely Positive Maps (TPCPMs). This class of maps includes \emph{any} possible process that is compatible with the laws of quantum theory i.e., any process obeying assumption $\Laws$, including for instance electromagnetic radiation as in Bernstein's example.

\smallskip

\textit{\textbf{Information-Disturbance Theorem.}\footnote{The theorem may be phrased with respect to different measures of approximation (see also Ftn.~\ref{ftn_approximation}). For a precise quantitative version that is well-suited for our purposes, we refer to~\cite{kretschmann_information-disturbance_2008}. }
  Let  $\mathcal{C}_{A \to B}$ be a TPCPM that describes a desired reversible quantum computation with input $A$ and output $B$. Let $\mathcal{D}_{A \to B E}$ be a TPCPM that describes a physical device with input $A$ and output $B$ as well as leakage to the environment $E$. Then
\begin{align} \label{eq_informationdisturbance}
  \mathcal{D}_{A \to B} \approx  \mathcal{C}_{A \to B}  
    \quad \implies \quad \mathcal{D}_{A \to E} \approx \mathcal{S}_{A \to E} \ ,
\end{align}
where $\mathcal{S}_{A \to E}$ is a perfectly scrambling channel from $A$ to $E$, i.e., $E$ is uncorrelated to $A$.}

\smallskip

Hence, if the device $\mathcal D_{A\to BE}$ realizes the desired computation $\mathcal C_{A\to B}$ with little error, then nothing about the identity of the input could have leaked to any environmental degrees of freedom. Conversely, if the environment $E$ does gain some information about the input, then some disturbance inevitably occurs. In the above example, the phase of the superposition of the input state is not maintained by the \textsc{cnot} gate.\footnote{In this example, the desired operation $\mathcal{C}_{A \to B}$ is the identity map on $A= B= Q$, and the operation of the physical device $\mathcal{D}_{A \to B E}$ is the map from $A = Q$ to $B E = Q Q'$ induced by the \textsc{cnot} gate when $Q'$ is initialized to $\ket{b_0}$. Since $\mathcal{D}_{A \to E}$ leaks the classical bit in $A$ to $E$ it has large distance to the scrambling channel $\mathcal{S}_{A \to E}$. The theorem thus implies that $\mathcal{D}_{A \to B}$ must have large distance to the identity map on $Q$, which is indeed the case.}

In order to process information secretly, it is therefore sufficient to ensure that the premise of~\eqref{eq_informationdisturbance} holds, i.e., that the device   suffers only little disturbance during the computational process. 
Low disturbance is precisely the goal of fault-tolerant quantum computation.
Threshold theorems ensure that if our control of physical systems is accurate enough so that we can realize a set $\mathfrak{G}$ of basic operations at a noise level below a certain threshold, then arbitrarily-low logical error rates can be achieved by a sufficient amount of error correction (see \cite{gottesman_introduction_2010} and references therein). While the theorems differ in their precise assumptions and claims, they usually have the following generic structure.

\smallskip

\emph{\textbf{Fault-Tolerance Threshold Theorem.}
  Let $\mathfrak{G}$ be a set of gates, described by TPCPMs, and denote by $\NoiseThreshold$ the assumption that any element from a given universal gate set\footnote{See~\cite{nielsen_quantum_2010} for examples for universal gate sets, which are usually finite.} is realized by a gate from $\mathfrak{G}$ with error probability not exceeding a given threshold $\theta$ and that there is negligible  correlation between noise on different gates. Then for any desired TPCPM $\mathcal{C}_{A \to B}$  there exists a device $\mathcal{D}_{A \to B}$ consisting only of elements from $\mathfrak{G}$  such that
\begin{align} \label{eq_threshold}
  \NoiseThreshold \quad \implies \quad \mathcal{D}_{A \to B} \approx \mathcal{C}_{A \to B} \ .
\end{align}
}

The two theorems above refer to the processing of general quantum information, whereas in a QKD setup Alice and Bob merely require classical computation to process their measured data. But this is no difficulty. Let $\mathcal{\tilde{C}}_{A \to \tilde{B}}$ be a TPCPM that describes the local operations that need to be carried out according to a given $\AbsPro$. By virtue of Stinespring's dilation theorem (see~\cite[Theorem 2.22]{watrous_theory_2018} or \cite[Section 8.2.4]{nielsen_quantum_2010}), it is always possible to regard $\mathcal{\tilde{C}}_{A \to \tilde{B}}$ as part of a reversible quantum computation $\mathcal{C}_{A \to B}$, as required by the Information-Disturbance Theorem stated above. Let now $\mathfrak{G}$ be a set of gates compatible with $\Laws$, and suppose that it satisfies $\NoiseThreshold$. Then the Fault-Tolerance Threshold Theorem implies that $\AbsPro$ can be turned into a $\PhysPro$ whose TPCPM description $\mathcal{D}_{A \to B E}$ satisfies $\mathcal{D}_{A \to B} \approx  \mathcal{C}_{A \to B}$.\footnote{For Alice and Bob to be able to run the desired protocol they need ``protected labs'' in the (weak) sense that an adversary cannot significantly influence the outcome of an experiment done entirely within these labs. This assumption should be understood as a part of $\PhysPro$.}  But this means, according to the Information-Distur\-bance Theorem, that $\PhysPro$ leaks virtually no information to the environment. Combining this with~\eqref{eq_securityproofabstract}, we can conclude that\footnote{\label{ftn_approximation}The approximations in~\eqref{eq_informationdisturbance} and~\eqref{eq_threshold} can be understood as equalities with a small error probability~$\varepsilon'$. The latter can be included in the security parameter~$\varepsilon$ that is implicit to the security claim $\Secrecy$, as mentioned in Ftn.~\ref{ftn_epsilon}.}
\begin{align} \label{eq_NoiseThresholdImpl}
  \Laws \wedge \PhysPro \wedge \NoiseThreshold \quad \implies \quad \Secrecy \ .
\end{align}

This statement may not yet be satisfactory for practical applications. Since the gate set $\mathfrak{G}$ appearing in the derivation sketched above is universal for quantum computation, nominally Alice and Bob require a quantum computer to run $\PhysPro$ --- although according to $\AbsPro$ the data they need to process is classical. 
However, for a purely classical computation a smaller set of gates  $\mathfrak G_{\text{c}}$ universal to classical computation is sufficient. This has been shown in \cite{lacerda_classical_2019} by reduction to the quantum setting\footnote{It is of course still assumed that the classical computer, on the physical layer, obeys the laws of quantum theory. In \cite{lacerda_classical_2019} the reduction also requires that $\mathfrak{G}_{\text{c}}$ includes a gate that generates random bits that do not immediately leak to the eavesdropper.}, and can again be seen from the simple example above: 
To guard classical information encoded into a quantum system against leakage it is sufficient to ensure that the encoding allows for the correction of phase errors. 
When viewed as actions on the classical information, such an encoding is an instance of a masking countermeasure mentioned in \cite[Section 6]{bernstein_is_2018}.\footnote{However, in contrast to the countermeasures  proposed in~\cite{bernstein_is_2018}, which merely increase Eve's cost for accessing the secret information, the argument presented here implies that Eve gains no information (except with small probability~$\varepsilon$, see Ftn.~\ref{ftn_epsilon}).}
The connection between leakage and phase errors is entirely similar to the manner in which privacy amplification in QKD (which is also applied to classical data) is understood as phase error correction in, e.g., the QKD security proof of Shor and Preskill~\cite{shor_simple_2000}.  

Let us summarize our conclusions so far. We have used the Information-Disturbance Theorem and the Fault-Tole\-rance Threshold Theorem --- both of which are well-established statements of quantum information theory --- to infer~\eqref{eq_NoiseThresholdImpl}. However, to obtain claim~\eqref{eq_securityproofphysical}, we still need to show that there exist constructions for physical gate sets $\mathfrak{G}$ such that 
\begin{align} \label{eq_PhysicalRealisability}
  \Laws \quad \implies \quad  \NoiseThreshold \ .
\end{align}

By now there are many concrete constructions of physical gate sets which satisfy this implication. 
They differ in their choices of physical information carriers~\cite{humble_quantum_2019} and hence also in the specific physical laws they rely on. Take for example the proposal of Cirac and Zoller~\cite{CiracZoller} for the \textsc{cnot} gate. Here the information is carried by cold trapped ions and manipulated by electromagnetic pulses, and the relevant physical laws are therefore those of Quantum Electrodynamics. Although the construction is aimed at  quantum computing, it serves our purposes: It proves the physical realisability of gates for which~\eqref{eq_PhysicalRealisability} holds. 

We have thus established claim~\eqref{eq_securityproofphysical}, which in turn implies that Bernstein's claim~\eqref{eq_leakage} is false. As a byproduct, this also justifies~\eqref{eq_quantumclassical}: Classical laws cannot be obtained from quantum laws if~\eqref{eq_classicalleakage} holds but ~\eqref{eq_leakage} does not.

\subsection{What physics assumptions are necessary?}

As noted in~\cite{bernstein_is_2018}, the physics assumptions are not usually stated explicitly in security proofs. However, examining the general line of reasoning leading to~\eqref{eq_securityproofphysical}, as presented above, one can identify the elements that must be included in $\Laws$. The first part, which uses the Information-Disturbance and the Fault-Tolerance Threshold Theorem and leads to \eqref{eq_NoiseThresholdImpl}, is phrased entirely within the abstract formalism of quantum information theory. It thus only depends on the general structure of the quantum-mechanical state space and the unitarity of time evolution. The second part of the argument, which leads to~\eqref{eq_PhysicalRealisability}, involves the concrete construction of gate sets~$\mathfrak{G}$, and thus relies on specific physical laws. For instance, the Cirac-Zoller construction from~\cite{CiracZoller} mentioned above is based on quantum electrodynamic models of lasers and ion traps. 

One may still question whether real devices are accurately described by the particular physics models. This corresponds to questioning the accuracy of assumption~$\Laws$, not the security claim~\eqref{eq_securityproofphysical} itself. It could be that standard quantum theory is too broad, and that future refinements of $\Laws$ may rule out various TPCPMs and, for instance, render the implication~\eqref{eq_PhysicalRealisability} moot. 
This issue can only be decided by experiments. 

Most directly, experiments may establish the realisability of  gate sets $\mathfrak{G}$ that satisfy $\NoiseThreshold$. 
While it follows that, if no such gate sets exist, then secrecy of the local processing cannot be ensured (at least not via the argument presented here), it also follows in that case that the current approach to realize scalable quantum computers cannot work, either.\footnote{This is indeed a point of contention for the possibility of fault-tolerant quantum computation for a small number of researchers~\cite{alicki_critique_2013,kalai_argument_2020}. 
The objection is that noise processes will necessarily involve non-negligible coupling between qubits.}
Though a scalable quantum computer has not yet been constructed, its building blocks have already been realized.  For instance, individual Cirac-Zoller gates were implemented in 2003~\cite{schmidt-kaler_realization_2003}. More recently, multi-qubit gates at a much larger scale have been demonstrated (so-called ``quantum supremacy'')~\cite{arute_quantum_2019}. 
Experimental efforts in error correction have also recently taken great strides~\cite{negnevitsky_repeated_2018,andersen_repeated_2020}. 
While these experiments do not yet settle the question whether realistic gate sets $\mathfrak{G}$ that satisfy $\NoiseThreshold$ can be built, we take them as promising evidence in this direction.%
\footnote{If and when reliable quantum information processors of sufficient size for processing Alice and Bob's secret data can be built, they could even themselves test whether assumption $\NoiseThreshold$ holds.} 

\subsection{Beyond quantum physics.}

Finally, let us comment on Bernstein's suggestion that the holographic principle, like Maxwell's equations, implies that leakage is inevitable. 
This takes us into the realm beyond quantum physics, as the holographic principle is not contained in the quantum formalism used above for $\Laws$. 
Here we take issue with Bernstein's argument for a different reason. 
The security statement is a mathematical statement, and the extent to which we say physical law underlies the statement is the extent to which we believe the mathematical formalism of the theory used in the security statement captures physical phenomena. 
In our view, current theories beyond the usual quantum formalism, such as quantum gravity or string theory, are still in a rather preliminary stage. In particular, they do not describe cryptographic hardware in a  mathematically precise manner that would allow us to formulate a security statement.
That is, we do not know how to formulate \eqref{eq_securityproofphysical} for $\Laws$ beyond the standard quantum formalism. 

Nevertheless, \eqref{eq_leakage} could hold true if one replaces $\Laws$ by the laws of a future theory beyond quantum physics.  It could certainly happen that, according to such a theory (one including gravity, for instance), leakage is inevitable. 
One does not even have to go very far to find an example of such a breach with standard quantum theory: 
If the hidden variables of the de Broglie-Bohm theory of quantum mechanics (see~\cite{bricmont_making_2016} for an overview) were to become accessible after all, then leakage would again go unnoticed, as in the case of classical physics.\footnote{In this case the assumption of completeness of quantum theory would be violated; cf.\ Ftn.~\ref{ftn_correctcomplete}.}   

\section{Conclusions}
Quantum physics has the remarkable feature that it imposes fundamental constraints on the accessibility of information. 
In particular, the uncertainty principle requires any effort to gain information about a physical system to be accompanied by disturbance to that system.  
This feature lies at the heart of quantum cryptography, as it enables the honest parties to infer the eavesdropper's information gain by observing the attendant disturbance. 
As shown in this note, the same feature of quantum physics also establishes that classical information can be stored and processed in a leakage-free manner. 
Without the latter, claims of secrecy would indeed be inconsistent with the laws of physics, and~\eqref{eq_securityproof} would be vacuous.

\subsection*{Acknowledgements}
\hspace{0.5em}We acknowledge financial support from the Swiss National Science Foundation via the National Center for Competence in Research for Quantum Science and Technology (QSIT) and via the QuantERA project No.\ 20QT21\_187724, and from the Air Force Office of Scientific Research (AFOSR) via grant FA9550-19-1-0202.

\printbibliography[heading=bibintoc,title={\normalsize References}]

\end{document}